\documentclass{article}
\usepackage[T1]{fontenc}
\usepackage{microtype}
\usepackage{graphicx}
\usepackage{subfigure}
\usepackage{booktabs}
\usepackage{amsmath,amssymb}
\usepackage{xcolor}
\usepackage{hyperref}

\usepackage[accepted]{mlsys2025}
\newcommand{\sys}{budgeter}
\mlsystitlerunning{Bursts That Outlast Themselves}

\begin{document}
\newcommand{\HoneFwd}{9}\newcommand{\HoneRev}{0}\newcommand{\HoneBoth}{2}\newcommand{\HoneNeither}{1}\newcommand{\HoneN}{12}\newcommand{\HoneP}{0.00391}

\twocolumn[
\mlsystitle{Bursts That Outlast Themselves: Persistent Degradation in Prefix-Cached Agent Serving}
\begin{mlsysauthorlist}
\mlsysauthor{Wen Huang}{msft}
\mlsysauthor{Jack Liu}{ind}
\end{mlsysauthorlist}
\mlsysaffiliation{msft}{Microsoft}
\mlsysaffiliation{ind}{Independent Researcher}
\mlsyscorrespondingauthor{Wen Huang}{wenhuang@microsoft.com}
\mlsyskeywords{LLM serving, prefix caching, metastability, agents}
\vskip 0.3in
\begin{abstract}
LLM agents drive inference engines with closed-loop sessions whose context grows every turn, and serving them
economically depends on prefix caching: a hit costs an incremental prefill of about a thousand tokens, a miss
recomputes a 10--40K-token context. We show that this makes prefix-cached agent serving \emph{metastable}. With an
identical base workload, a few minutes of extra arrivals leaves a vLLM server in a degraded regime---a
$\sim$10\% hit ratio, about half the throughput, and a backlog that grows for the remaining 70 minutes---while its
no-burst twin stays healthy (\HoneFwd{} of \HoneN{} pairs; none in the reverse direction). The regime appears at
loads the cache-warm server sustains, is absent without prefix caching, and also appears with Qwen3-32B on SWE-bench
agent trace shapes and on SGLang, which tolerates larger triggers. Slower service keeps more sessions in the system,
their working set outgrows the KV pool, and every turn becomes a recomputation; a first-order workload-balance
condition, calibrated independently, brackets when the regime can drain. Getting out is the harder problem:
throttling new admissions recovered 1 of 7 matched collapsed runs, while suspending admitted sessions against a
capacity-based limit recovered all of them within minutes. A CPU offload tier delays the collapse but does not prevent it. A budgeter that bounds the resident KV demand of active
sessions to the pool, with thresholds that need no per-workload tuning, prevented the failure in every run we
tested.

\end{abstract}
]
\makeatletter\renewcommand{\Notice@String}{Preprint.}\makeatother
\printAffiliationsAndNotice{}
\chead{\small\bf Bursts That Outlast Themselves}

\section{Introduction}
\label{sec:intro}

Coding assistants, browser agents and other tool-using LLM agents now account for a large and growing share of
inference traffic. An agent session is a closed loop: the model emits a tool call, the tool runs, its output is
appended to the conversation, and the grown context goes back to the model---tens to hundreds of times per task.
Serving this traffic economically depends on \emph{prefix caching}~\cite{zheng2024sglang,vllmapc}. If the
previous turn's KV cache is still resident, a turn costs an incremental prefill of a few hundred to a few thousand
tokens; if it was evicted, the engine recomputes the whole context, which for an agent is 10--40$\times$ more work.
Production coding-agent traces show KV cache hit ratios averaging 90\% within a user-initiated agent loop~\cite{liu2026copilot}.

\begin{figure}[t]
\centering
\includegraphics[width=\columnwidth]{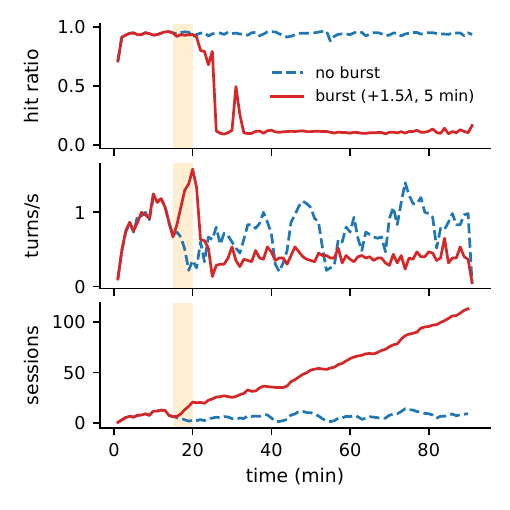}
\vspace{-1em}
\caption{A burst that outlasts itself. Same model, same server, same base arrivals; the red run additionally receives a
5-minute injected burst (shaded). Its hit ratio and throughput never recover after the burst ends, while its blue
no-burst twin stays healthy.}
\label{fig:teaser}
\end{figure}

This paper is about what happens when that hit ratio is lost. Figure~\ref{fig:teaser} shows two runs of a vLLM
server with \emph{the same sessions arriving at the same times}. In one, we inject a few minutes of extra arrivals.
During the burst, the hit ratio dips, as one would expect. What one would not expect is what happens next: after
the burst ends and the arrival rate returns to a level the server was handling comfortably, the hit ratio does not
come back. It settles near 10\%, throughput falls below half, and the number of sessions in the system grows for
the remaining hour. The twin run without the burst never leaves its healthy state. The extra load is long gone;
the backlog stays.

This is a \emph{metastable failure} in the sense of \citet{bronson2021metastable} and \citet{huang2022metastable}: a trigger pushes the system into a degraded state, and a sustaining effect keeps it
there after the trigger is removed. The sustaining effect here is specific to stateful LLM serving. Agent sessions
are closed-loop, so the number of sessions concurrently in the system grows with response time (Little's law). Every
active session needs its prefix resident to hit. When service slows, more sessions overlap, their combined context
outgrows the KV pool, LRU evicts each session's prefix before its next turn, every turn becomes a full recomputation,
and service slows further.

Several recent systems have observed this feedback loop as concurrency rises: CONCUR~\cite{chen2026concur} in offline
agentic batch inference, MISA-T~\cite{hong2026misat} in RL rollout serving, VAMP~\cite{kim2026vamp} as a
``prefix-cache cliff'' in multi-turn MoE serving, and KAIROS~\cite{yuan2026kairos}, which notes that a short overload
can leave an instance persistently inefficient. They keep the system out of the loop with cache-aware concurrency
control (including agent pause/resume in CONCUR), KV-demand-based admission with hysteresis, or HBM repartitioning. We do not claim the loop or session admission as new. Our contribution is about
\emph{dynamics}: the loop can be entered from a load the server sustains, it does not exit on its own when the load
returns, and leaving it calls for a different control action than staying out of it. Concretely:

\begin{itemize}
\item \textbf{The phenomenon, controlled.} Using paired runs that share every base session and differ only in an
injected burst, we show history-dependent, persistent degradation on vLLM with Qwen3-8B (\HoneFwd{} of \HoneN{} pairs, none in the reverse direction); its
absence without prefix caching at matched sub-capacity load; and its reproduction with Qwen3-32B on the engine's
full KV pool replaying SWE-bench agent trace shapes, and on SGLang, which tolerates larger triggers
(\S\ref{sec:phenomenon}).
\item \textbf{Why it stays.} In the degraded regime nearly every turn misses and the server runs at its no-cache
capacity. A first-order workload-balance condition, with per-miss cost calibrated independently on cache-disabled
runs, explains when the regime cannot drain (\S\ref{sec:model}).
\item \textbf{Getting out.} Once collapsed, throttling admissions recovered 2 of 8 runs after 21--23 minutes and
latency-driven AIMD (which also suspends sessions) 5 of 7 after 14--25 minutes, while suspending admitted sessions
at turn boundaries against a limit derived from the KV pool restored throughput and a 0.93--0.95 hit ratio in every
collapsed run within 6 minutes. A budgeter that bounds the resident KV
demand of active sessions to a fraction of the pool does so without workload-specific tuning
(\S\ref{sec:controller}, \S\ref{sec:eval}).
\end{itemize}

\section{Background: Agent Sessions and Prefix Caching}
\label{sec:background}

\paragraph{Closed-loop agent sessions.}
An agent session alternates model calls and tool executions. Each call sends the full conversation---system
prompt, task, and every prior model output and tool observation---and the next call is issued only after the
tool returns. Context therefore grows by the model output plus the tool output on every turn.
In our analysis of public coding-agent traces (SWE-bench runs of 2026 frontier models), a session makes a median
of 42 calls, appends $\sim$350 tokens per turn, and reaches 16--40K tokens; on the TraceLab dataset of day-to-day
Claude Code and Codex usage~\cite{zhu2026tracelab} we find a median starting context of 19K tokens and a median
per-call context of 136K. Tool gaps are short (median 0.1--0.2\,s in these two datasets) with a heavy tail.
Unlike open-loop chat traffic, the offered load of a session population depends on the server: a turn cannot be
issued until the previous one has been served, so the number of sessions concurrently in the system is
in steady state $N = \lambda \cdot T \cdot (Z + R)$ by Little's law, where $\lambda$ is the session arrival rate,
$T$ the number of turns per session, $Z$ the mean think/tool time and $R$ the mean per-turn response time. While a
backlog is growing this is not an identity, but the direction holds: slower turns mean more concurrent sessions.

\paragraph{Prefix caching.}
Engines such as vLLM~\cite{kwon2023vllm,vllmapc} and SGLang~\cite{zheng2024sglang} keep the KV cache of
completed requests in a paged pool (indexed by prefix hash in vLLM, by a radix tree in SGLang) and reuse matching
blocks for later requests. For agents,
the previous turn's KV is a prefix of the next turn, so a hit reduces the prefill of turn $k$ from $L_k$ tokens to
the $\sim$$g$ newly appended tokens. Cached blocks are evicted (LRU) when running requests need space; the pool
is shared between the KV of running requests and cached prefixes of idle sessions.

\paragraph{Why misses are so expensive.}
On an MI300X serving Qwen3-8B, we measure cold prefill at 10--13K tokens/s for 20--36K-token contexts,
versus incremental prefill on a cached prefix at 5--16K new tokens/s and decode at 20--190 tokens/s per request
depending on batch size and context. A mid-session turn (context 21.6K) costs $\sim$2\,s of prefill if it misses and
$\sim$0.15\,s if it hits---a ratio that grows linearly with session length.

\section{A Burst That Outlasts Itself}
\label{sec:phenomenon}

\paragraph{Setup.}
We run vLLM 0.20 (ROCm build \texttt{0.20.2rc1.dev253}) with automatic prefix caching on single AMD MI300X GPUs,
one engine per GPU, BF16 weights and KV cache, default scheduler (chunked prefill, FCFS), 40K maximum context,
greedy decoding with a fixed output length per turn; every run starts from a freshly started engine, and the arms
of a pair or seed share the same base arrival stream (32B no-burst baselines: same seed only) and run on separate GPUs, mostly concurrently. The main configuration serves Qwen3-8B (144\,KiB of KV per token) with the KV pool fixed at
400K tokens (59\,GB), close to the $\approx$365K-token pool Qwen3-32B gets on the same GPU; \S\ref{sec:phen-32b} removes this restriction.
A closed-loop driver issues sessions as a Poisson process of rate $\lambda$. Each synthetic session starts with
a 2K-token system prompt shared by all sessions plus a 4K-token task unique to the session, runs 25 turns of 300
output tokens, appends 1K tool tokens per turn, and waits an exponential think time (mean 5\,s) between turns;
contexts grow from 6K to 37K tokens. Token contents are random, so the only reuse is exact prefix reuse within a
session and of the shared system prompt.

To isolate history dependence we use \emph{paired} runs. Both runs of a pair share one base arrival stream:
every base session arrives at the same time with the same identity and shape. The burst run additionally receives
an independent stream of extra sessions (rate $1.5\lambda$ for 5 minutes, i.e. 2.5$\times$ total arrivals) starting
at minute 15. Everything after the burst is therefore identical in offered base load. Runs last 90 minutes.

We report three endpoints: the token-weighted prefix hit ratio of completed turns, $\sum_i c_i/\sum_i p_i$ over
turns $i$ completing in a window, with $c_i$ the engine-reported cached prompt tokens and $p_i$ the prompt length, and
turn throughput (the engine); the number of sessions in the system, i.e.\ arrived and not finished, whether
running, thinking, parked or waiting for admission (the work); and the completion of arrival cohorts (the
users). A run is
\emph{persistently degraded} if every 5-minute window (non-overlapping, aligned to the end of the trigger) from its first degraded window (hit ratio $<0.5$) after the
trigger to the end of the run is degraded.

\begin{table*}[t]
\centering\footnotesize
\caption{Experiments. Times are minutes from the start of a run; runs last 30--120 minutes. Synthetic sessions: 25
turns, 6K$\to$37K tokens (session-mix shift: 12 turns, 3K/22K start); trace shapes: mini-SWE-agent on SWE-bench, truncated at 38K tokens.}
\label{tab:matrix}
\setlength{\tabcolsep}{2.5pt}
\resizebox{\textwidth}{!}{%
\begin{tabular}{llllllr}
\toprule
Experiment & Model / engine & KV pool & Base $\lambda$ (/s) & Trigger & Control & Runs \\
\midrule
Paired history dependence (\S\ref{sec:phenomenon}) & 8B / vLLM & 400K & 0.035 & +1.5$\lambda$, 5 min @15 & -- & 12 pairs \\
No-cache control & 8B / vLLM, no cache & 400K & 0.0086 & +16 sessions, 5 min @15 & -- & 4 pairs + 4 \\
Flush trigger & 8B / vLLM & 400K & 0.04 & cache flush @10 & none / \sys{} & 5 + 3 \\
Trace shapes & 32B / vLLM & $\approx$365K & 0.025 & 2$\times$, 4 min @10 & none / \sys{} / cap & 4 each \\
Second engine & 8B / SGLang & 400K & 0.04 & 2$\times$ 2 min; +1.5$\lambda$ 5--10 min; flush & -- & 4+6+3, 4 pairs \\
Recovery (\S\ref{sec:eval}) & 8B / vLLM & 400K & 0.04 & 2$\times$, 4 min @10 & 5 arms @20 & 16 seeds \\
Recovery, trace shapes & 32B / vLLM & $\approx$365K & 0.03; 0.025 & 2.5$\times$ / 3$\times$, 5 min & 5 arms; 3 arms @25 & 4; 3 seeds \\
Session-mix shift & 8B / vLLM & 400K & 0.075 & short$\to$long sessions @30 & 7 arms & 2 seeds \\
CPU KV offload (128\,GiB) & 8B / vLLM & 400K & 0.035; 0.04 & as paired; 2$\times$, 4 min @10 & --; none / \sys{} @20 & 6 pairs; 3 seeds \\
Drain condition (\S\ref{sec:model}) & 8B / vLLM & 400K & 0.04$\to$0.006--0.018 & +1.5$\lambda$, 5 min @10 & -- & 6 + 2 partial \\
\bottomrule
\end{tabular}}
\end{table*}

\paragraph{Same load, different fate.}
At $\lambda{=}0.035$ sessions/s---about 70\% of the cache-warm capacity, estimated as $\approx$0.05
sessions/s from the up to 1.27 turns/s (25 turns per session) the engine sustains at a 0.93--0.95 hit ratio after recovery---most no-burst runs stay healthy for the
full 90 minutes (hit ratio 0.94--0.95, $\sim$10 sessions in the system). Figure~\ref{fig:teaser}
shows a typical pair. In the burst run, the hit ratio falls during the burst and does not return: it settles at
$\approx$0.1, turn throughput falls to $\sim$0.4 turns/s, and the backlog grows past 100 sessions by the end of the
run. Table~\ref{tab:h1} lists all completed pairs: in \HoneFwd{} of \HoneN{} only the burst run degraded persistently,
in \HoneRev{} only the no-burst run did, \HoneBoth{} degraded in both arms and \HoneNeither{} in neither (two-sided sign
test over discordant pairs, $p=\HoneP$). The healthy regime is not immune to fluctuations: occasionally a no-burst
run also slid into the degraded regime on its own after tens of minutes and, like the burst runs, never left it.
A trigger makes entering the regime far more likely; nothing we observed makes leaving it likely.

\begin{table}[t]
\centering\small
\caption{Paired runs with an identical base workload: token-weighted hit ratio over the last 10 minutes and mean
number of sessions in the system over the last 5 minutes of the 90-minute run. $^*$persistently degraded from its
first degraded window after the burst to the end of the run.}
\label{tab:h1}
\begin{tabular}{rcccc}
\toprule
 & \multicolumn{2}{c}{No burst (twin)} & \multicolumn{2}{c}{Burst} \\
Seed & Hit & Sessions & Hit & Sessions \\
\midrule
501 & 0.95 & 8 & 0.12$^*$ & 109 \\
502 & 0.95 & 8 & 0.11$^*$ & 142 \\
503 & 0.95 & 8 & 0.12$^*$ & 144 \\
504 & 0.95 & 9 & 0.12$^*$ & 142 \\
505 & 0.95 & 6 & 0.95 & 5 \\
506 & 0.95 & 9 & 0.10$^*$ & 143 \\
507 & 0.13$^*$ & 137 & 0.11$^*$ & 175 \\
508 & 0.95 & 5 & 0.11$^*$ & 159 \\
509 & 0.95 & 6 & 0.12$^*$ & 144 \\
510 & 0.12$^*$ & 124 & 0.12$^*$ & 147 \\
511 & 0.95 & 6 & 0.12$^*$ & 129 \\
512 & 0.95 & 7 & 0.11$^*$ & 112 \\
\bottomrule
\end{tabular}

\end{table}

\paragraph{Not just overload.}
Three controls separate the persistent regime from ordinary overload. (i) Ten of the twelve no-burst twins run at the same base
load without degrading; the other two degraded spontaneously, 25 and 30 minutes into the run. This is the metastable signature, not a sign
of overload: a system with a persistent degraded regime leaves the good state on rare large fluctuations of its own
arrivals, while 10 of 10 runs at $\lambda{=}0.04$ without a burst stayed healthy for 30 minutes in an earlier sweep. (ii) With prefix caching disabled, at about 80\% of the no-cache engine's capacity, we inject the same absolute
burst as above ($\approx$16 extra sessions over 5 minutes). The backlog rises to 17--43 sessions and then falls:
over the last 30 minutes it shrinks in every run (or is already at its pre-burst level), and two of four runs
end at their twins' level (3 and 1 unfinished sessions); the other two are still draining at a cache-less engine's slow pace. With a smaller, rate-proportional burst, both arms of all four
pairs end with the same small backlog (0--5 unfinished sessions). Without a cache there is no second regime to fall into---only work to
catch up on. Nor is it the burst's own sessions still being served: in 7 of 11 degraded burst runs all injected
sessions had finished, yet the hit ratio stayed at 0.10--0.12. (iii) With a session cap that keeps the active population below what the KV pool
holds, the same bursts never cause persistent degradation (0/10 runs in an earlier matrix).

\paragraph{Triggers other than bursts.}
A prefix-cache flush with preemption of running requests---what a replica restart or weight update does---at
$\lambda{=}0.04$ and without any extra load left 1 of 5 runs persistently degraded; the run that collapsed had the
largest number of sessions arriving while the cache was being rebuilt. Whether a given trigger tips a run is not determined by
its size alone: in the paired experiment, bursts injecting 11 and 13 extra sessions collapsed their runs while one
injecting 17 did not, so the outcome also depends on the state of the sessions already in the system when the
trigger arrives.

\begin{figure}[t]
\centering
\includegraphics[width=\columnwidth]{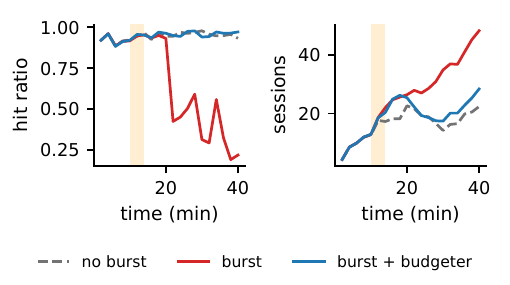}
\vspace{-1em}
\caption{Qwen3-32B, automatically sized KV pool, SWE-bench trace shapes, $\lambda{=}0.025$ (same seed): a 2$\times$
burst from minute 10 to 14, the same burst with the budgeter, and no burst.}
\label{fig:32b}
\end{figure}

\paragraph{A larger model on its full KV pool, with agent trace shapes.}
\label{sec:phen-32b}
We replay whole-session shapes---initial context, per-turn output, per-turn context growth and inter-turn gaps---of
200 trajectories sampled from public mini-SWE-agent leaderboard runs on SWE-bench Verified by six 2026 models
(token counts and timestamps as reported by the providers; median 42 calls per session, 16K-token median context per call; 26\% of sessions exceed 38K tokens and are
truncated there, removing 14\% of turns), with a unique task suffix per session (engine limit 40K), against Qwen3-32B with its KV pool sized automatically at 80\% GPU-memory utilization
($\approx$365K tokens, no explicit cap). At $\lambda{=}0.025$, a 4-minute 2$\times$ burst left 1 of 4 runs persistently degraded and a second partially
degraded at the end of a 40-minute window (end hit ratio 0.34 and 0.81 vs. 0.95--0.97 for all four no-burst runs,
Figure~\ref{fig:32b}), but a longer no-burst run degraded on its own after about 55 minutes, so 32B is weaker evidence of
history dependence than 8B; at
$\lambda{=}0.035$ the server ran healthily for about 15 minutes and then degraded without any burst. At $\lambda{=}0.03$ the 32B
configuration did not stay healthy even without a trigger: all 4 runs with a 5-minute burst degraded, and so did all 4
no-burst runs, after 25--35 minutes (hit ratio 0.12--0.17 at about 79 minutes). We therefore treat
$\lambda{=}0.03$ as beyond the healthy side for this configuration and use it only to study recovery
(\S\ref{sec:eval}); history dependence on 32B rests on the $\lambda{=}0.025$ runs.

\begin{figure}[t]
\centering
\includegraphics[width=\columnwidth]{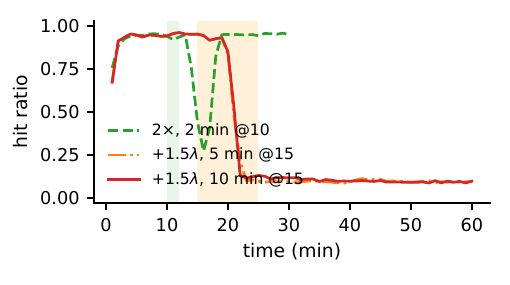}
\vspace{-1em}
\caption{SGLang at $\lambda{=}0.04$ absorbs a 2$\times$, 2-minute burst (from minute 10) but not an extra
$1.5\lambda$ for 5 or 10 minutes (from minute 15).}
\label{fig:sglang}
\end{figure}

\paragraph{A second engine.}
SGLang~0.5.22 with RadixAttention, the same model, pool size and nominal workload parameters (an earlier driver version, so arrivals are not identical), absorbed in 4 of 4 runs the 2$\times$/2-minute burst
that collapses vLLM in 4 of 10: in two SGLang runs the hit ratio dipped (to 0.27--0.29 in 1-minute windows), in
the other two it barely moved, and all recovered within minutes. A larger
trigger did collapse it: an extra $1.5\lambda$ for 5 or 10 minutes left SGLang at a 0.09--0.11 hit ratio with a backlog
growing to 52--112 sessions at the end of the run, 35--40 minutes after the burst, in all 6 runs (3 seeds
$\times$ 2 burst lengths, Figure~\ref{fig:sglang}). SGLang refuses to flush its cache while requests run; once a
flush went through (after retrying for up to 19\,s), the server recovered in 3 of 3 runs, unlike vLLM, where a
flush with preemption collapsed 1 of 5. In paired runs with the design of the paired vLLM experiment above (identical base arrivals, +1.5$\lambda$ for 5 minutes) at $\lambda{=}0.04$, all 4 SGLang burst runs degraded
persistently (hit ratio 0.09--0.11) while all 4 no-burst twins stayed healthy (0.93--0.95).
The failure is not specific to one engine; engine design moves its threshold.

\section{Why the Backlog Stays}
\label{sec:model}

\paragraph{Anatomy of the degraded regime.}
In the persistently degraded 8B runs, essentially all turns miss: over the last 10 minutes of the paired burst runs,
99.6--100\% of completed turns had less than half of their prompt cached, and the token-weighted hit ratio is
0.10--0.12, which is just the shared system prompt. The engine is busy, not stalled: 16--32 requests run
concurrently, KV usage sits at 83--100\%, and preemptions remain rare (23--37 in a 90-minute run), so the lost
throughput is recomputation rather than thrashing on preemption. The active population, meanwhile, far exceeds what the pool can hold at the sessions'
average context, $N^* = C/\bar L \approx 18$, and keeps growing; each session's prefix is evicted by its peers before
its next turn---an analogue of working-set thrashing~\cite{denning1968thrashing}. Turn throughput in this regime (0.35--0.46
turns/s) is that of the same engine with prefix caching disabled (0.33--0.43 turns/s at comparable load).

\begin{figure}[t]
\centering
\includegraphics[width=\columnwidth]{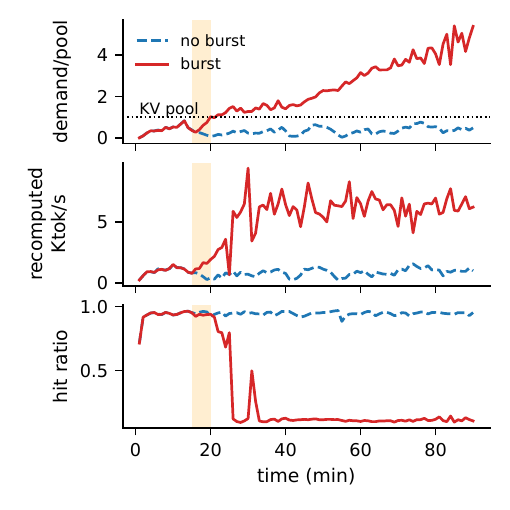}
\vspace{-1em}
\caption{The loop in one pair (seed 501): resident demand relative to the KV pool, recomputed prompt tokens, and hit
ratio, per minute. The burst (shaded) pushes demand past the pool; the twin never crosses it.}
\label{fig:mechanism}
\end{figure}

Figure~\ref{fig:mechanism} shows the loop forming in one pair. Before the burst, the resident demand of the
sessions in the system (their number times the mean context of their turns) is about 0.6 of the KV pool. The burst
pushes it past the pool (1.4$\times$ at minute 25); from then on the hit ratio collapses and recomputed prompt
tokens rise from about 1.1K to 6.4K per second, while demand keeps climbing to 4.8$\times$ the pool over the last 5 minutes. The
no-burst twin peaks at 0.83 of the pool and never crosses it. We infer, rather than measure, that the misses come
from prefixes evicted between turns; the engine does not export per-block eviction events.

\paragraph{The loop.}
Three properties of agent serving close the loop. (1) \emph{Backlog is working set}: in a closed loop the number of
active sessions is, in steady state, $N = \lambda T (Z + R)$, so a slower response time $R$ directly enlarges the
set of prefixes that must stay resident. (2) \emph{The miss penalty grows with session age}: a miss recomputes the whole context, which
grows every turn, so as sessions in the degraded regime age, each miss costs more. (3) \emph{Running requests and
the cache share one pool}: the more requests run, the less room remains for idle sessions' prefixes.

\paragraph{When can the degraded regime drain?}
If every turn misses, serving one session costs $S = \sum_{k=0}^{T-1} c(L_k)$ engine-seconds, where $c(L)$ is the
cost of a cold turn with context $L$. The degraded regime can drain only if arrivals bring less work than the engine
completes, i.e. if $\lambda < 1/S$. We measure $c(L)$ independently, on runs with prefix caching disabled: from 48 complete two-minute windows in which
at least four requests were running, we regress window length on the number of completed turns and their total
prompt tokens, giving $c(L) = -0.06\,\text{s} + 0.146\,\text{ms} \times L$ (tokens; the intercept is a fit artifact), hence $S \approx 78$\,s and
$1/S \approx 0.013$ sessions/s for the synthetic workload---well below the loads at which we observe collapse.
(Fitting instead on all-miss windows of degraded cache-enabled runs gives $c(L) = -0.71\,\text{s} + 0.187\,\text{ms}\times L$ and $1/S{=}0.012$.)  This is a necessary condition, not a prediction of the boundary. When we collapse the server and then lower the
arrival rate, the estimated remaining cold work (fit over the whole post-change interval at 10-minute samples)
falls by about 900 engine-seconds per hour at $\lambda{=}0.006$ (one seed), stays roughly flat at
$\lambda{=}0.010$ ($-67$ to $+132$/h over three seeds), and grows at $\lambda{=}0.014$ (+604 and +663/h, two
seeds) and $0.018$ (one seed, sessions in the system 53 to 100 over the hour observed). No run's
cache re-formed while observed (2 hours; 80 minutes for two partial runs). The break-even thus lies near
0.010--0.014, bracketing the model's 0.013; the drift estimates depend on the fitting window, so we report the bracket.

\paragraph{The healthy side.}
The healthy regime exists as long as the cache-warm population fits, $\lambda T (Z + R_0)\,\bar L \lesssim C$, and
the cache-warm engine keeps up. Between the two conditions the outcome depends on history.  A larger KV pool widens the healthy side but leaves the drain
condition unchanged, so it raises the trigger size needed rather than eliminating the failure, and rare
combinations---a burst, a restart, an unlucky arrival cluster---rather than average load determine when a server
collapses. We do not model the escape time from the healthy regime quantitatively; our attempt to do so with a
Poisson fluctuation model was too pessimistic by an order of magnitude.

\section{Getting Out}
\label{sec:controller}

A collapsed server has two problems: too many sessions want the cache at once, and every one of them keeps issuing
turns. Admission control addresses only new arrivals. The sessions already in the system keep cycling through a
cache that cannot hold them; each must finish its remaining turns---all misses---before the population falls below
$N^*$, which takes tens of minutes at the degraded regime's throughput. Recovery is faster if the gateway also
\emph{suspends} some admitted sessions: a parked session does not issue turns, its prefix is allowed to be evicted,
and the remaining sessions start hitting again, which restores throughput and drains the queue.

\paragraph{One actuator, several limits.}
We implement all controllers at the gateway (no engine changes) with the same actuator, so that they differ only in
how they set the limit. Before each turn of a session, the gateway decides whether the session may proceed. A
session that is not admitted, or that would push the active population over the limit, is \emph{parked} at this turn
boundary and waits in a FIFO queue. Parked sessions are re-admitted before any new session, one at a time as the
limit allows; a re-admitted session usually pays one miss, since its prefix was likely evicted while it waited. The limits we compare are:
\begin{itemize}
\item \textbf{\sys{}}: the resident KV demand $D=\sum_{s\in\mathcal{A}} (L_s + o_s)$ of the set $\mathcal{A}$ of
  admitted sessions (context $L_s$ plus output budget $o_s$) is held in a hysteresis band: sessions are admitted while
  $D \le 0.75C$ and parked when $D > 0.95C$, so $D$ can exceed $0.75C$ as admitted sessions grow, and briefly
  exceed $0.95C$ between turn boundaries. $C$ is the engine's KV-pool size, which the engine reports at startup.
\item \textbf{Session cap + park}: admit up to $K$ sessions (park above $K_{hi}$), with $K$ set from the
  per-session demand measured on the healthy workload ($K{=}14$, $K_{hi}{=}17$ in \S\ref{sec:eval}).
\item \textbf{Adaptive cap + park}: $K = \lfloor 0.75C/\hat d\rfloor$ (park above $\lfloor 0.95C/\hat d\rfloor$),
  where $\hat d$ is an EWMA ($\alpha{=}0.02$ per turn) of the demand $L_s+o_s$ observed at turn boundaries.
\item \textbf{Latency AIMD + park}: $K$ adapted every 10\,s by additive increase / multiplicative decrease
  (halving when the median turn latency exceeds 15\,s, initial $K{=}32$); a simplified latency-driven proxy for
  concurrency control in the spirit of CONCUR~\cite{chen2026concur}, not a reimplementation of it.
\item \textbf{Admission-only cap}: new sessions wait while $K{=}16$ sessions are active; admitted sessions are never parked.
\end{itemize}

\paragraph{Algorithm.} At each turn boundary of session $s$ the gateway (1) updates $s$'s demand to its current
context plus output budget; (2) if $s$ is admitted and the limit's park condition holds (for the \sys{},
$D > 0.95C$), parks $s$---the session that has just reached a boundary, so no in-flight request is interrupted---and
appends it to the FIFO queue; otherwise $s$ proceeds; (3) whenever a turn completes or a session ends, re-admits
waiting sessions from the head of the queue, parked sessions before new ones, while the admission condition holds
(for the \sys{}, $D + d_s \le 0.75C$). In-flight requests count toward $D$ with their full context and output
budget. A parked session keeps its place in the queue and is re-admitted as a whole; it can be parked again at a later
boundary if the band is exceeded again. The
thresholds 0.75/0.95 were fixed before the experiments in \S\ref{sec:eval} and not tuned per workload.

\paragraph{Why tokens, not sessions.}
A session-count limit is a token limit divided by an assumed per-session demand. Agent sessions differ by an order
of magnitude in context length, and the mix shifts---interactive sessions during the day, long autonomous runs at
night---so a fixed count is either too loose when sessions are long or too tight when they are short. The \sys{}
needs only $C$ and its two thresholds, and adapts to the mix automatically. $D$ is a conservative proxy: it
double-counts shared prefixes and is updated at turn boundaries, so it bounds the active working set rather than the
physical residency of the cache.

\section{Evaluation of Recovery and Prevention}
\label{sec:eval}

\begin{figure}[t]
\centering
\includegraphics[width=\columnwidth]{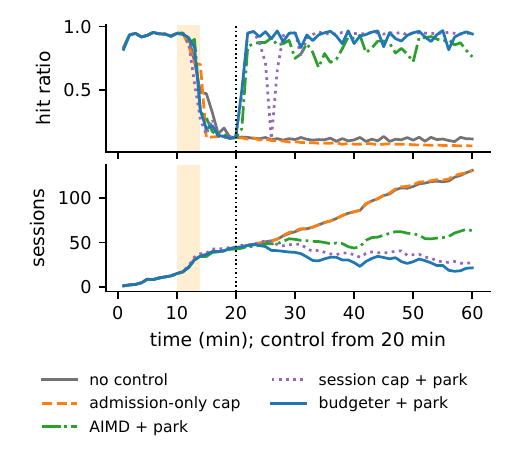}
\vspace{-1em}
\caption{Recovery with no control and four controllers enabled at minute 20, one seed (identical arrivals; this seed was
collapsed at minute 20 in every arm).}
\label{fig:recovery}
\end{figure}

\begin{figure}[t]
\centering
\includegraphics[width=\columnwidth]{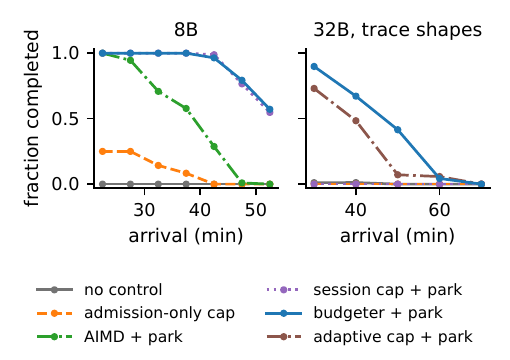}
\vspace{-1em}
\caption{Fraction of each arrival cohort (5-minute cohorts for 8B, 10-minute for 32B) that completed by the end of
the run (minute 60 for 8B, 90 for 32B), averaged over the runs collapsed when control began. Later cohorts have less
time to complete.}
\label{fig:cohort}
\end{figure}

\paragraph{Recovery after collapse.}
We collapse the server first (Qwen3-8B, 400K-token pool, $\lambda{=}0.04$, a 2$\times$ burst for 4 minutes at minute
10) and switch on each controller at minute 20, running to minute 60. All arms of a seed share the same arrivals.
Of 16 seeds, 7--8 per arm were collapsed when control began; Table~\ref{tab:recovery} reports those.
Restricting to the seven seeds collapsed in every arm, which start control from matched states (29--48 sessions
in the system and 421--831 remaining turns, each arm within 6\% of its seed's mean), gives the same ordering: the
\sys{} recovered all seven in under 5 minutes, the session cap all seven within 6 minutes, latency
AIMD five of seven after 14--25 minutes, the admission-only cap one of seven, and no control none. On these seeds the \sys{} and the
session cap each recovered 7 of 7 versus 1 of 7 for the admission-only cap (two-sided exact paired $p{=}0.031$ for
each comparison); across the 8B, 32B and threshold-sensitivity recovery experiments the \sys{} recovered 17 of 17
collapsed runs.
The admission-only cap recovered 2 of 8 runs, 21--23 minutes later; four stayed below a 0.1 hit ratio to the end.
Latency AIMD with parking recovered 5 of 7 after 14--25 minutes. Both limits that park against the pool's capacity
recovered every collapsed run: the \sys{} within the first 5-minute measurement window after control began, and the
session cap within at most 6 minutes (recovery time is the start of the first 5-minute window, stepped by 1 minute,
from which every later window has a hit ratio $\ge$0.8, requiring at least 10 minutes of follow-up; we therefore report times below 5 minutes as ``$<$5''), with
a 0.93--0.95 hit
ratio and every session arriving in the 20 minutes after control began completing (Figures~\ref{fig:recovery} and~\ref{fig:cohort}).

\begin{table*}[t]
\centering\small
\caption{Recovery when control starts at minute 20 (the 4-minute burst ran from minute 10 to 14), over runs that
were collapsed at that point; denominators differ because arms share arrivals but not engine state.
Time: from when control starts to the start of the first 5-minute window from which every later window has a hit
ratio $\ge$0.8; ``$<$5'' means within the first window. Cohort: fraction of sessions arriving in the
next 20 minutes that completed.}
\label{tab:recovery}
\begin{tabular}{lcccc}
\toprule
Control (from 20 min) & Recovered & Time to recover (min) & Hit (end) & Cohort done \\
\midrule
None & 0/8 & -- & 0.12--0.13 & 0\% \\
Admission-only cap & 2/8 & 21--23 & 0.07--0.93 & 0--92\% \\
Latency AIMD + park & 5/7 & 14--25 & 0.84--0.89 & 50--100\% \\
Session cap + park & 7/7 & $\le$6 & 0.95 & 100\% \\
\sys{} + park & 7/7 & $<$5 & 0.93--0.95 & 100\% \\
\bottomrule
\end{tabular}

\end{table*}

\begin{figure}[t]
\centering
\includegraphics[width=\columnwidth]{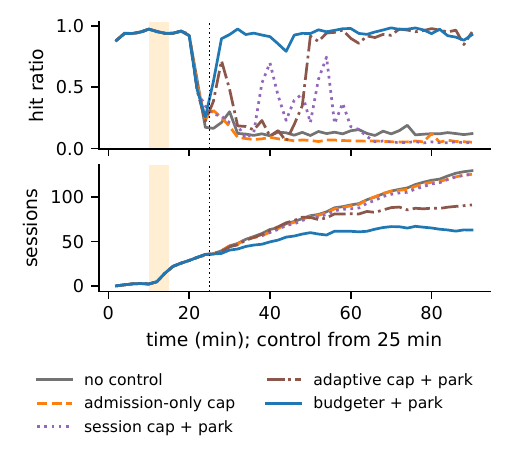}
\vspace{-1em}
\caption{Qwen3-32B, full KV pool, SWE-bench trace shapes: recovery when control starts at minute 25 (one seed).}
\label{fig:h3}
\end{figure}

On this homogeneous workload the \sys{} and a session cap tuned to the measured per-session demand perform alike; the
difference is that the cap had to be tuned. On heterogeneous sessions the difference matters: with Qwen3-32B,
its full KV pool and SWE-bench trace shapes, a session cap set from a lighter load never recovered a collapsed
server, and neither did an admission-only cap, in any of the three seeds collapsed in every arm when control began (0/3
each, end hit ratio 0.05--0.08 for the admission-only cap); an EWMA-adaptive cap recovered within 21 minutes,
and the \sys{} recovered in under 5 minutes in all three (hit ratio 0.92--0.97,
Figure~\ref{fig:h3}).

At $\lambda{=}0.025$, where the 32B server stayed cache-efficient without a trigger over the 40-minute runs of
\S\ref{sec:phenomenon}, we repeat the recovery
experiment with a stronger trigger (3$\times$ arrivals for 5 minutes) and 100-minute runs, reporting all three seeds.
Without control every run degraded and stayed degraded (hit ratio 0.12--0.13; 142--175 sessions in the system over
the last 5 minutes). With control from minute 25, the \sys{} and the adaptive cap restored the hit ratio in every collapsed run
(0.92--0.97 and 0.86--0.93 over the last 10 minutes) and completed 54--96\% and 83--97\% of the sessions arriving
in the following 20 minutes, versus 0--7\% without control. The queue did not drain within the run, however: with
either controller it kept growing slowly (4--8 sessions per 10 minutes over the last 30 minutes, averaging 73--101
sessions over the last 5 minutes), because over the long run this trace mix arrives slightly faster than the cache-warm 32B engine
completes it (1.5--1.6 arrivals vs.\ 0.7--1.3 completions per minute). The controllers restore the engine's
efficiency; they cannot create capacity that is not there, and draining the backlog needs additional capacity or
load shedding.

\paragraph{Threshold sensitivity.}
Changing the \sys{}'s thresholds from 0.75/0.95 to 0.65/0.85 or 0.85/0.98 of the pool changed neither the outcome
nor the timing on the 8B recovery runs we repeated: every run recovered in under 5 minutes, with
end hit ratios of 0.92--0.95, all sessions arriving in the following 20 minutes completed, and post-recovery
throughput similar to the default setting (1.01--1.25 vs.\ 1.06--1.27 turns/s; not a strictly matched comparison).

\begin{figure}[t]
\centering
\includegraphics[width=\columnwidth]{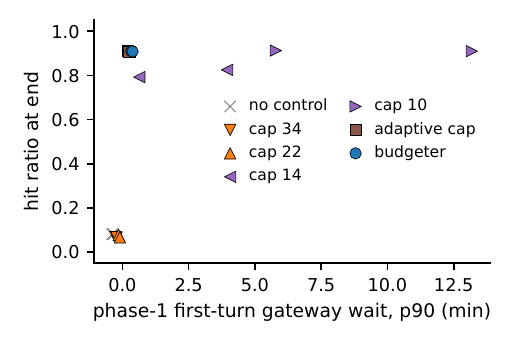}
\vspace{-1em}
\caption{Workload shift from short to long sessions. Each point is a run: loose fixed caps collapse after the shift
(bottom), tight ones make first-phase sessions wait (right); only the budgeter and the adaptive cap sit at the top
left. Two seeds per configuration; points offset by $\le$0.4 min for visibility.}
\label{fig:h4}
\end{figure}

\paragraph{When the session mix shifts.}
A count limit is only as good as its assumed per-session demand. We run Qwen3-8B (400K pool) at $\lambda{=}0.075$
with short interactive sessions (3K-token initial context, 15\,s think time) for 30 minutes, after which arriving
sessions are long autonomous ones (22K initial context, 5\,s think time)---no burst. Without control, the shift
alone drives the server into the degraded regime (2/2 seeds, hit ratio 0.08 at the end).
Figure~\ref{fig:h4} shows the trade-off a fixed session cap faces: caps of 22 or 34 sessions, comfortable for the
short sessions, are too loose for the long ones and collapse like the uncontrolled server (4/4); caps of 10 or 14
prevent collapse but over-throttle the short-session phase, holding first turns at the gateway for a p90 of 0.7--13
minutes. The \sys{} and the EWMA-adaptive cap avoid both failure modes (0/4 collapsed, no gateway wait in the first
phase, and every session arriving in the first 60 minutes completed by the end of the 80-minute run).

\paragraph{What users see.}
Queued sessions still have to be served (Figure~\ref{fig:cohort}). On 32B with trace shapes, the \sys{}
completed 65--92\% of the sessions arriving in the 20 minutes after control began by the end of the 90-minute run
(63--98 still in the system; completions stayed below arrivals over the last 30 minutes), an adaptive cap 35--95\%,
and no-control, admission-only and fixed-cap runs 0--3\%. Parking delays some sessions to raise throughput
for all. On the seven 8B seeds collapsed in every arm, sessions arriving
in the 20 minutes after control began took a median of 11.6 minutes from arrival to completion with the \sys{}
(p90 16.9) and 12.3 (p90 15.6) with the session cap, versus 3.8 minutes in never-collapsed uncontrolled runs; with latency AIMD,
75 of 349 had not finished by the end and the rest took a median of 20.6 minutes; with the admission-only cap 317
of 349 had not finished, and without control none finished.

\paragraph{Prevention.}
With the \sys{} active from the start, no run collapsed under any trigger we applied: 0/5 with a 2$\times$, 2-minute burst, 0/3
with a cache flush, and 0/4 with Qwen3-32B and SWE-bench trace shapes (where one uncontrolled run and one run under a session
cap tuned on a lighter load ended partially degraded). Its throughput in the healthy regime was within run-to-run variation of
the uncontrolled server.

\paragraph{CPU KV offloading.}
A host tier does not remove the regime. With vLLM's native CPU offloading (128\,GiB, $\approx$0.9M tokens) in the
paired design, 4 of 6 burst runs ended persistently degraded (hit ratio 0.12, 116--136 sessions in the system) while
all 6 twins stayed healthy; in the recovery design all 3 uncontrolled runs collapsed late, while the \sys{}, enabled at minute 20 before
collapse, kept all 3 healthy. Offloading only delays the collapse (persistent degradation from minute 40--60
instead of 20--25), consistent with the growing population outgrowing the host tier as well. Faster offload paths would raise the threshold; we tested the \sys{} here only for prevention.

\section{Related Work}
\label{sec:related}

\paragraph{Metastable failures.}
\citet{bronson2021metastable} define metastable failures as
states in which a sustaining effect keeps a system degraded after its trigger
has passed, and \citet{huang2022metastable} show they underlie
many major cloud outages.
Such failures have been modeled with CTMCs~\cite{isaacs2025analyzing,alvaro2025formal}, traced to
destabilizing cycles among stable components~\cite{farahbakhsh2026characterizing}, and modeled as
queues in replicated storage~\cite{habibi2024msf}. \citet{huang2022metastable} also show capacity-degradation amplification, including a cache whose hit ratio does
not recover after the trigger is removed. We identify such a sustaining effect specific to stateful LLM serving: the
prefix-cache hit ratio, coupled to the population of closed-loop sessions, and
show that suspending work leaves the degraded state far more reliably than shedding new load.

\begin{table*}[t]
\centering\footnotesize
\caption{What prior agent-serving work establishes about cache-induced collapse. ``Same-load history dependence'':
identical subsequent load, different outcome depending on a past trigger. ``Recovery of admitted'': measured
recovery once collapsed, with control over already-admitted sessions. ``--'': not demonstrated in the cited
evaluation; CONCUR has agent pause/resume, but does not measure recovery from an established collapse.}
\label{tab:related}
\setlength{\tabcolsep}{6pt}
\begin{tabular}{lcccc}
\toprule
 & Loop & Same-load & Persistence & Recovery of \\
 & observed & history dependence & after trigger & admitted sessions \\
\midrule
CONCUR~\cite{chen2026concur} & \checkmark & -- & -- & -- \\
KAIROS~\cite{yuan2026kairos} & \checkmark & -- & argued & -- \\
MISA-T~\cite{hong2026misat} & \checkmark & -- & -- & -- \\
VAMP~\cite{kim2026vamp} & \checkmark & -- & -- & -- \\
Metronome~\cite{meng2026metronome} & KV-pool exhaustion & -- & -- & -- \\
This paper & \checkmark & \checkmark & \checkmark & \checkmark \\
\bottomrule
\end{tabular}
\end{table*}

\paragraph{Working sets and load control.}
Denning showed that a multiprogrammed system thrashes when the working sets of active programs exceed memory, and
that the remedy is load control: deactivate programs until the rest fit~\cite{denning1968working,denning1968thrashing}.
The \sys{} is that remedy applied at the gateway, and we claim no novelty for it. What is specific to agent serving is that
sessions are closed loops: at a fixed session arrival rate, cache misses lengthen every
session's lifetime and so enlarge the competing working set, which makes degradation history-dependent at a load the
server otherwise sustains; admission throttling alone exits it only slowly or not at all (\S\ref{sec:eval}). Our contribution is
establishing that regime experimentally and showing which form of load control gets out of it.

\paragraph{Concurrent work.}
After our submission, \citet{lu2026basin} independently reported the same failure on prefill--decode-disaggregated
SGLang (H200/H100, Qwen3.6-35B-A3B): a 10-minute surge cuts goodput by about 88\% for the following hour, sustained by
the same population--cache feedback. They analyze it with a fluid model and simulations across deployments, and
propose Basin, which orders and budgets the admission of new sessions; admitted sessions run to completion, and
service recovered once the burst sessions were terminated. Our study is complementary: a single colocated engine on
vLLM and SGLang, a cache-disabled control, a measured drain condition, and a comparison of controllers that act on
already-admitted sessions after collapse, where parking recovers runs that admission throttling alone does not.

\paragraph{Cache-induced collapse in agentic serving.}
Several systems observe that prefix-cache efficiency collapses as agentic
concurrency grows. CONCUR~\cite{chen2026concur} reports ``middle-phase
thrashing'' in offline agentic batch inference. MISA-T~\cite{hong2026misat}
describes the feedback loop between admitted sessions, the resident working set and cold prefill. VAMP~\cite{kim2026vamp} calls the resulting threshold a
``prefix-cache cliff.'' KAIROS~\cite{yuan2026kairos} argues that even a short overload can push an
instance into a persistently inefficient region and avoids it with context-occupancy admission and hysteresis, but
does not test it with a removed trigger or measure recovery once collapsed. EfficientAgent~\cite{shao2026efficientagent} applies
Denning's load control to host-tier KV offloading. Table~\ref{tab:related} summarizes selected systems; we do not claim the
loop or session admission as new, only the dynamics around them. Metronome's~\cite{meng2026metronome} ``metastable'' collapse is KV-pool exhaustion in speech serving.
Eviction-policy work (RR-Evict~\cite{pan2026rrevict}, LRU studies on
production traces~\cite{liu2026fancy}), production residency
characterization~\cite{arya2026kvresidency}, working-set capacity
planning~\cite{li2026kvset}, and queueing-aware placement~\cite{zou2026vesta}
replay recorded sessions rather than ask whether a burst leaves lasting degradation; KV-centric stores~\cite{qin2025mooncake} make the same
reuse-or-recompute choice. Here, the trace responds to misses.

\paragraph{Models of LLM serving.}
\citet{yao2026multiturn} derive a mean-field, Che-style~\cite{che2002hierarchical}
hit ratio for multi-turn conversations under LRU and show that it falls with
arrival rate. Their analysis assumes exogenous inter-turn times, excluding the latency feedback that
sustains the degraded regime here. \citet{ao2026sic} find multiple equilibria and limit
cycles caused by KV growth during decoding, but with prefix caching disabled
and saturated input. \citet{nie2026queueing} and \citet{dong2026flow} derive stability conditions under KV-memory
constraints for single-turn requests, leaving multi-turn caching open.

\section{Discussion}
\label{sec:discussion}

\paragraph{For operators.}
The degraded regime is easy to misdiagnose: GPU utilization stays high, latency alarms fire, and waiting or
throttling new traffic rarely brings the cache back. The prefix hit ratio of agent traffic and the number of
sessions in the system relative to what the KV pool holds identify it directly; parking sessions at turn boundaries
was the fastest and most reliable remedy we tested. Restarts and weight updates are triggers in their own right;
ramping admission while the cache rebuilds avoids the collapse. The queued work still has to drain afterwards (10--30
minutes on our 8B setup).

\paragraph{Engine-side alternatives.}
SGLang tolerated triggers that collapsed vLLM, so engine design moves the threshold. We did not compare
session-aware eviction or scheduling that favors requests with a resident prefix; both act on the quantity the
\sys{} bounds and could raise the threshold, but once the active population's demand exceeds the pool some
sessions must wait, and the choice is only where.

\paragraph{Threats to validity.}
Synthetic sessions use random tokens; the trace replay preserves lengths and timing but not content, truncates at
38K tokens, and does not execute agents. We use one GPU type, one replica and finite windows (persistence for 70
minutes after the burst, and 100 minutes after lowering the arrival rate), with no routing, autoscaling, client
retries or production evidence. Production contexts are longer (median per-call above 100K on TraceLab), which for a
fixed pool lowers $N^*$ and raises the miss penalty; we have not measured it. On 32B the backlog never drained, so
our evidence on queue stability is from 8B.

\section{Conclusion}
\label{sec:conclusion}
Prefix caching turns the KV pool of an agent-serving engine into a working set whose size is set by the backlog.
A transient trigger can push the server into a regime where the backlog keeps the cache from re-forming and the
missing cache keeps the backlog growing, at loads the server otherwise sustains. Suspending admitted sessions, not only refusing new ones,
gets a server out of that regime quickly and reliably in our experiments; a budgeter bounding the resident
demand of active sessions to the pool prevented it in our runs and restored the cache when it occurred.

\bibliography{refs}
\bibliographystyle{mlsys2025}

\section*{AI Usage Disclosure}
We used AI assistants (Anthropic Claude and OpenAI Codex) in this work. The authors proposed the research question
and the overall idea, and led the experiments: they designed the experimental plan, chose the models, workloads and
hardware, decided which follow-up experiments to run, and judged which results support which claims. AI assistants
helped check the topic against prior work by searching for and summarizing related papers, pull requests and model
releases. Under the authors' direction, they helped refine the experimental plan and assisted with code, experiment
runs, figures and tables, and writing. The authors reviewed all AI-assisted code and text, checked the reported
numbers and citations, and take full responsibility for the content.

\end{document}